\documentclass[conference]{IEEEtran}

\usepackage{mathrsfs}

\usepackage{amssymb}

\usepackage{amsthm}

\usepackage[mathcal]{euscript}

\usepackage{psfrag,calc,url,bm}

\usepackage{cite}

\usepackage{graphicx}

\usepackage{psfrag}

\usepackage{subfigure}

\usepackage{url}

\usepackage{stfloats}

\usepackage{amsmath}

\usepackage{float}

\usepackage{algorithmic}

\usepackage[ruled,linesnumbered,vlined]{algorithm2e}

\usepackage{boxedminipage}

\usepackage{color}

\usepackage{setspace}

\usepackage{soul}

\begin{document}

\title{Optimal Online Transmission Policy in Wireless Powered Networks with Urgency-aware Age of Information}

\author{Yang Lu$^{*,\Diamond}$, Ke Xiong$^{*,\Diamond}$, Pingyi Fan$^\dag$, Zhangdui Zhong$^{\natural,\S}$, and Khaled Ben Letaief$^\ddag$\\
\small
$^*$School of Computer and Information Technology, Beijing Jiaotong University, Beijing 100044, China \\
$^\Diamond$ Beijing Key Laboratory of Traffic Data Analysis and Mining, Beijing Jiaotong University, Beijing 100044, China\\
$^\dag$Department of Electronic Engineering, Tsinghua University,
Beijing 100084, China\\
$^\natural$State Key Lab of Rail Traffic Control and Safety, Beijing Jiaotong University, Beijing 100044, China\\
$^\S$Beijing Engineering Research Center of High-speed Railway Broadband Mobile Communications,\\
 Beijing Jiaotong University, Beijing 100044, China \\
 $^\ddag$ The Hong Kong University of Science and Technology, Hong Kong\\
E-mail: kxiong@bjtu.edu.cn, fpy@tsinghua.edu.cn}

\maketitle

\begin{abstract}
\textcolor[rgb]{0,0,0}{This paper investigates the age of information (AoI) for a radio frequency (RF) energy harvesting (EH) enabled network, where a sensor first scavenges energy from a wireless power station and then transmits the collected status update to a sink node. To capture the thirst for the fresh update becoming more and more urgent as time elapsing, \emph{urgency-aware AoI (U-AoI)} is defined, which increases exponentially with time between two received updates. Due to EH, some waiting time is required at the sensor before transmitting the \textcolor[rgb]{0,0,0}{status} update. To find the optimal transmission policy, an optimization problem is formulated to minimize the long-term average U-AoI  under constraint of energy causality. As the problem is non-convex and with no known solution, a two-layer algorithm is presented to solve it,  where the outer loop is designed based on Dinklebach's method, and in the inner loop, a semi-closed-form expression  of the optimal waiting time policy is derived based on Karush-Kuhn-Tucker (KKT) optimality conditions. Numerical results shows that our proposed optimal transmission policy outperforms the the zero time waiting policy and equal time waiting policy in terms of long-term average U-AoI, especially when the networks are non-congested. It is also observed that in order to achieve the lower U-AoI, the sensor should  transmit the next update without waiting when the network is congested while should  wait a moment before transmitting the next update when the network is non-congested. Additionally, it also shows that the system U-AoI first decreases and then keep unchanged with the increments of EH circuit's saturation level and the energy outage probability.}


\end{abstract}

\section{Introduction}

With development of wireless sensor networks (WSN) and Internet of Things (IoTs), more and more sensors will be deployed in networks to monitor  the status of environment, devices and living creatures, such as temperature, speed and heart rate. Then, the status updates are sent to a sink node for some applications such as smart city, industrial control and human \textcolor[rgb]{0,0,0}{health} detection. Some traditional metrics, e.g., delay \cite{new1,new2} and throughput\cite{new3,new4}, are widely adopted to evaluate the system performance {of wireless networks}. However, for some real-time applications {including} \textcolor[rgb]{0,0,0}{smart} drive, keeping the collected status {data} ``fresh" becomes a vital concern, and the traditional metrics cannot reflect such kind of requirements. To capture the ``freshness" of the status update, {a new metric referred to as} age of information (AoI) has drawn great attentions  {recently}. AoI, which was first defined in \cite{first_aoi}, denotes the amount of time that elapsed since the moment the freshest received update was generated. {The AoI oriented  network design is very different from traditional throughput and delay oriented network designs}. For a network aiming to maximize the throughput, the status updates may be generated and transmitted with a very high frequency, which may cause networks congestion and high delay among delivered status updates. For a network aiming to minimize the delay, the status updates may be generated and transmitted with a very low frequency, which may cause the last received status update becomes stale.  {\textcolor[rgb]{0,0,0}{However}, to keep information fresh  (with minimal AoI), the status updates should not  be generated at neither a high frequency nor a low frequency.}

 {So far, the system AoI performance has been studied in some existing works for various wireless systems}. For instance, in \cite{first_aoi} and \cite{aoi3}, the AoI was analyzed for a single-source single-server network and a multiple-source $M/M/1$ system, respectively. In \cite{aoi},  {it was pointed out} that increasing the number of servers at the sink node can reduce the average AoI but may cause waste of network resources. In \cite{AoI1}, an optimal transmission was designed to minimize long-term average AoI, which showed that zero time waiting policy does not always minimize the average AoI.

 {In the mentioned works \textcolor[rgb]{0,0,0}{above}, the sensors were supposed to be with fixed power supply. However, in practice,} it is with a tremendous economical drawback to charge the sensors with cables  {or batteries} due to expensive cost of installing and maintaining conventional battery recharging operation manually, especially in the hard-to-reach areas. As the sensors are usually ultra-low power, it is more efficient to use wireless power to charge them \cite{EH}. Compared with natural energy source, such as wind, solar, geothermal and hydropower, employing a dedicate power station to transfer energy is more {controllable and} reliable \cite{SWIPT1,SWIPT2,SWIPT3}. In particular, the sensor  {is equipped with a small} energy harvesting (EH) circuit to convert the received radio frequency (RF) signals  into direct current (DC) power. {That is, the wireless powered} sensors need to scavenge energy from the power station at first, and then,  {may use the harvested energy to} transmit status updates to the sink node.


\textcolor[rgb]{0,0,0}{So far, a few works have started  investigat\textcolor[rgb]{0,0,0}{ing} AoI in EH-enabled networks. In \cite{AoI2},  the update submission policy was optimized with a fixed update rate. In \cite{aoieh}, optimal online status update policies for an EH-enabled sensor were proposed with various battery sizes. In \cite{aoieh2}, the average AoI was analyzed for wireless powered networks in low SNR region.  {However, in these works, }, the energy arrivals were  {described as} to occur as a point process.  {But in practice, with wireless power transfer}, the sensor has to accumulate energy over a period of time to charge itself.   {Thus, the point process model is not suitable.} Moreover, in existing works, the AoI  was regarded to be linearly  {increased} with time  between the moments of two received updates, which cannot reflect the thirst for the fresh update that becomes more and more urgent as time elapsing  in many applications.}

 {To fill this gap, in this paper, we consider a scenario where a sink node collects status updates from a sensor charged by a wireless power station.} A waiting time is inserted before transmitting next status update such that the sensor could scavenge enough energy to transmit \textcolor[rgb]{0,0,0}{it}. To capture the thirst for the fresh update becoming more and more urgent as time elapsing,  {we adopt the \emph{urgency-aware AoI (U-AoI)} which exponentially increases with time  between the moments of two received updates to characterize the user experienced AoI  performance of the system. To minimize the system average U-AoI}, we formulate an optimization problem under the constraint of the energy causality. As the considered problem is non-convex  and with no known solution,  {we develop a two-layer algorithm to solve it,  where the outer loop is designed based on Dinklebach's method, and in the inner loop, a semi-closed-form expression  for the optimal waiting time policy is derived based on Karush-Kuhn-Tucker (KKT) optimality conditions.} Numerical results shows that the proposed transmission policy outperforms the zero time waiting policy and the equal time waiting policy in terms of long-term average U-AoI, especially when the networks are not congested. It is also observed that the sensor should  transmit  next update without waiting when the network is congested while should  wait a moment before transmitting next update when the network is not congested, in order to achieve the lower U-AoI. Besides, it is also shown that the system U-AoI first decreases and then keep unchanged with the increments of the  charging power level and the EH outage probability.

This paper is organized as follows. In Section II, the system model is introduced and the considered problem is formulated. In Section III, a status update policy is given. Some simulation results are presented in Section IV and this paper concludes in Section V.

\section{System Model and Problem Formulation}\label{systemmodel}

\subsection{System Model}

\begin{figure}[t]
\includegraphics[width=0.5\textwidth]{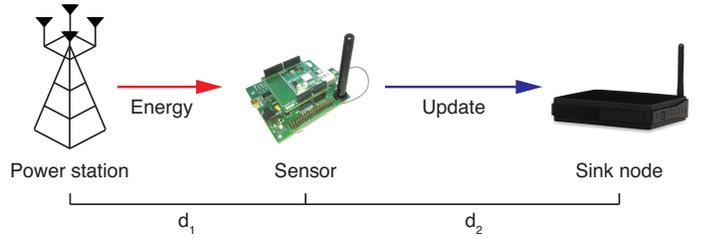}
 \caption{System model.}
 \label{Sys}
\end{figure}

\begin{figure}[t]
\includegraphics[width=0.5\textwidth]{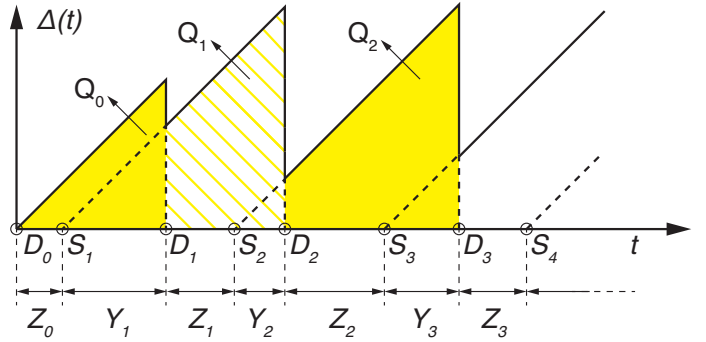}
 \caption{Evolution of the AoI $\Delta\left(t \right)$ at the sink node in terms of $t$.}
 \label{AoI}
\end{figure}

Consider an EH-enabled networks system as shown in Figure \ref{Sys}, where a power station provides energy to a sensor via wireless power transfer  and the sensor is in charge of continuously monitoring a sensing object, collecting status updates and transmitting the collected data to a sink node. For the sensor, it is energy-limited, so it may require a few seconds to scavenge energy from transmit signals by the power station  to fulfill the sensing and transmission task.

Let  ${P_{T}}$ denote the transmit power of the power station. The power carried in the received RF signals at the sensor can be given by
$$P_{R}\left( t \right) = \frac{{{{\left| {h_1\left( t \right)} \right|}^2}}}{{{d_1^\alpha }}}{P_{T}},$$
where $ {h_1\left( t \right)}$ denotes the channel coefficients of the links between the power station and the sensor at time $t$, $d_1$ denotes the distance, and $\alpha$ denotes the pass loss factor.

The non-linear EH model  \cite{Nonlinear1,Nonlinear2,Nonlinear3} is adopted to describe the harvested RF energy. Then, at time $t$, the output DC power (i.e., the harvested power) at the sensor is
 {\begin{flalign}
\Phi \left( {{P_R}\left( t \right)} \right) = \frac{{Me^{\left( {ab} \right)} - Me^{\left( { - a\left( {{P_R}\left( t \right) - b} \right)} \right)}}}{{e^{\left( {ab} \right)}\left( {1 + e^{ \left( { - a\left( {{P_R}\left( t \right) - b} \right)} \right)}} \right)}},\nonumber
\end{flalign}
where $M $ is a constant denoting the maximum output DC power, indicating the saturation limitation of the EH circuits. $a $ and $b $ are constants representing some properties of the EH system, e.g., the resistance, the capacitance and the circuit sensitivity.}

For the sink node, it keeps receiving the statues of the sensing object  with the received updates transmitted by the sensor.  For clarity, we use $i$ to represent index of the update with $i\in\left\{1,2,...\right\}$.

Suppose that the sensor generates and transmits the update\footnote{The package of update $i$ includes its generation time $S_i$. When it is delivered, a ACK including the delivered time $D_i$ will be fed back to the sensor from the sink node.} $i$ at time $S_i$ and the sink node receives the update $i$ at time $D_i$. Then, the transmission time for the update $i$ is $Y_i=D_i-S_i$, with $Y_i\ge0$.  {Let $C$ be the data size of each update}. Due to the channel uncertainty, the energy for transmitting one unit data may vary  versus time. Denote the transmit time as $T_c$, the bandwidth as $B$ and the transmit power  as $P_c$. Then we have that $$C = {T_c}B\log \left( {1 + \frac{{{{\left| {{h_2}\left( {{S_i}} \right)} \right|}^2}{P_c}}}{{d_2^\alpha B{n_0}}}} \right),$$
where $n_0$ denotes the noise spectral density, $d_2$ denotes the distance between the sensor and the sink node, and $ {h_2\left( {S_i} \right)}$ denotes the channel coefficient at time $S_i$. Then, at time ${S_i}$ the energy required for transmitting an update is $${\cal E}_i =P_cT_c= \frac{{\left( {{2^{\frac{C}{{{T_c}B}}}} - 1} \right)d_2^\alpha {T_c}B{n_0}}}{{{{\left| {{h_2}\left( {{S_i}} \right)} \right|}^2}}}.$$



\subsection{\textcolor[rgb]{0,0,0}{Average AoI and Average U-AoI}}

 {Due to \textcolor[rgb]{0,0,0}{variation} of network conditions, the transmission times $\left\{Y_1,Y_2,...,Y_i,...\right\}$ may vary from one update to another, which are treated as  $\emph{i.i.d}$ random variables.} When the update $i$ is received by the sink node, the sensor is notified to collect and transmit a new status update.   {As EH is employed, a period time of $Z_i$ may be  required to perform EH and detect and collect the sensing data}. That is, the amount of time to scavenge energy for transmitting the update $\left(i+1\right)$ is $\left(Y_i+Z_i\right)$. During this period, if the sensor has sufficient  energy, it may send the update to the sink node; otherwise, it  {has to accumulate energy}  at first. 

\begin{figure}[t]
\includegraphics[width=0.5\textwidth]{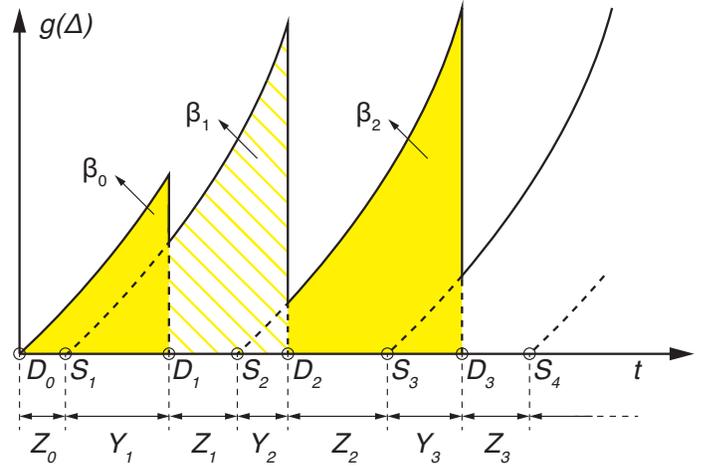}
 \caption{Evolution of the U-AoI $ g\left( \Delta\right)$ in terms of $t$}
 \label{AoI2}
\end{figure}

The time-stamped status updates should be as ``fresh" as possible at the sink node. At time $t$,  the generation time of the freshest update received at the sink node is
\begin{flalign}
U\left( t \right) = \max \left\{ {{S_i}\left| {{D_i} \le t} \right.} \right\},\label{freshest}
\end{flalign}
where,  {without loss of generality,} it is assumed that $S_0=0$.

\subsubsection{ {Traditional AoI}}

To measure the level of dissatisfaction for data (i.e., the updates) staleness,  traditional AoI is defined as
\begin{flalign}
\Delta \left( t \right) = t - U\left( t \right),\label{AoI_defination}
\end{flalign}
which denotes the amount of time that elapsed since the moment the freshest received update was generated.  With (\ref{AoI_defination}), AoI increases linearly with $t$ between the moments of two received updates and is with a downward jump when an update is received.

As illustrated in Figure \ref{AoI}, the integral of AoI over the time period of $\left[{D_0},{D_n}\right]$ is enclosed by the curve of $\Delta(t)$ and $t$-axis.  Thus, the average AoI of the system over $\left[{D_0},{D_n}\right]$ can be given by
\begin{flalign}
A\left( n \right) = \frac{{\int_{{D_0}}^{{D_n}} {\Delta \left( t \right)} dt}}{{{D_n} - {D_0}}}.
\end{flalign}
The numerator of  $A\left( n \right)$ can be calculated by  summing the ares of the  triangles and the trapezoids. Then,
$$\int_{{D_0}}^{{D_n}} {\Delta \left( t \right)} dt = \sum\nolimits_{i = 0}^{n-1} {{Q_i}}.$$
Further, the area of triangle or trapezoid $i$ between ${D_{i}}$ and ${D_{i+1}}$ can be calculated by   $$Q_i=\frac{1}{2}{\left( {{Z_i} + 2{Y_i} + {Y_{i + 1}}} \right)\left( {{Z_i} + {Y_{i + 1}}} \right)},~{\rm for}~i=0,1,...,n-1,$$
with $Y_0=0$. As a result, the expression of $A\left( n \right)$ can be given by
\begin{flalign}
A\left( n \right)  = \frac{{\sum\limits_{i = 0}^{n - 1} {\left( {{Z_i} + 2{Y_i} + {Y_{i + 1}}} \right)\left( {{Z_i} + {Y_{i + 1}}} \right)} }}{{2\left( {{D_n} - {D_0}} \right)}}.\nonumber
\end{flalign}
Actually, $A\left( n \right)$ reflects that the increment of data staleness keeps constant as time elapses, as the slopes of the lines associated with $\Delta(t)$ are constant.  {Thus, such a AoI model is a linear AoI model, which} is able to describe the AoI experience of some applications. But, for some real-time applications, such as online advertisement placement and online Web ranking, the linear AoI model may not be so efficient any more. Because in these applications, the dissatisfaction for date staleness  {(or the user's experience the thirst of the data freshness)} may grow more and more  quickly as time elapses.

\subsubsection{ {U-AoI}}

Thus, similar to \cite{AoI1}, we adopt an exponential function $g\left(\Delta \right):\left[ {0,\infty } \right) \to \left[ {0,\infty } \right)$ to denote the dissatisfaction for date staleness, which is given by
\begin{flalign}
 g\left(\Delta \right)=e^{a\Delta}-1,
 \end{flalign}
where $a\ge0$ is a pre-given constant to characterize the desire for data refreshing. The term ``-1" is to make  $g\left(\Delta \right)$ begin with zero. $g\left(\Delta \right)$ characterizes that the dissatisfaction for date staleness increases exponentially\footnote{Note that there exist other types of penalty functions such as $g\left(\Delta \right)=\ln\left({a\Delta+1}\right)$ which represents the logistic growth of dissatisfaction for date staleness over time.} as time elapses, which  means that with time elapsing, the thirst for the fresh update becomes more and more urgent. Therefore, we call $g(\Delta)$  as the \emph{U-AoI} in the sequel.

Similar to the traditional linear AoI, the average U-AoI  over $\left[{D_0},{D_n}\right]$ can be given by
\begin{flalign}
&{ A}_U\left( n \right) = \frac{{\int_{D_0}^{D_n} {g\left( \Delta  \right)} dt}}{D_n-{D_0}} = \frac{{\sum\nolimits_{i = 0}^{n-1} {\int_{{D_{i}}}^{{D_{i+1}}} {g\left( \Delta  \right)} dt} }}{D_n-{D_0}}. \nonumber
\end{flalign}

As mentioned previously, $Y_{i+1}=D_{i+1}-S_{i+1}$ and $S_{i+1}=D_{i}+Z_{i}$, we have that ${D_{i+1}} = {D_{i }} + {Z_{i }} + {Y_{i+1}}.$ Then, by defining the following auxiliary function for $t\in\left[D_{i},D_{i+1}\right]$, it is obtained that
\begin{flalign}
\beta_i &= \int_{{D_{i}}}^{{D_{i+1}}} {g\left( \Delta  \right)} dt \label{beta}\\ &= \int_{{D_{i }}}^{ {D_{i }} + {Z_{i }} + {Y_{i+1}}} {g\left( \Delta  \right)} dt.\nonumber
\end{flalign}
\textcolor[rgb]{0,0,0}{When $t\in\left[D_{i},D_{i+1}\right]$, the freshest update received at the sink node is the update $i$. Following (\ref{freshest}) and (\ref{AoI_defination}), we have
\begin{flalign}
{\Delta  = t - \underbrace {\left( {{D_i} - {Y_i}} \right)}_{{S_i}}}\label{beta:A}
\end{flalign}
By submitting (\ref{beta:A}) into (\ref{beta}), $\beta_i$ can be further expressed as
\begin{flalign}
\beta \left( {{Y_{i}},{Z_{i}},{Y_{i+1}}} \right) & = \int_{{Y_i}}^{{Y_i} + {Z_i} + {Y_{i + 1}}} {g\left( \tau  \right)} d\tau \label{beta:B}\\ &= \frac{1}{a}\left( {{e^{a\left( {{Y_i} + {Z_i} + {Y_{i + 1}}} \right)}} - {e^{a{Y_i}}}} \right) - {Z_i} - {Y_{i + 1}}.\nonumber
\end{flalign}}

As $D_n-{D_0} = \sum\nolimits_{i = 0}^{n-1} {\left( {{Y_{i+1}} + {Z_i}} \right)}  $, the long-term average U-AoI also can be calculated by
\begin{flalign}
{\mathbb E}\left[ {{A_U}\left( n \right)} \right] = \mathop {\lim }\limits_{n \to \infty } \frac{{{\mathbb E}\left[ {\sum\nolimits_{i = 1}^{n - 1} \beta  \left( {{Y_i},{Z_i},{Y_{i + 1}}} \right)} \right]}}{{{\mathbb E}\left[ {\sum\nolimits_{i = 0}^{n - 1} {\left( {{Y_{i + 1}} + {Z_i}} \right)} } \right]}}.
\end{flalign}


 \subsection{Problem Formulation}
For the considered wireless powered sensor network, our objective is to minimize the average U-AoI of the system under the constraint of the energy causality. To this end, an optimization problem aiming to find the optimal online schedule policy $\left\{ {{Z_0},{Z_1},{Z_2},...} \right\}$ is formulated by


\textcolor[rgb]{0,0,0}{\begin{subequations}
\begin{align}
\mathop {\min }\limits_{\left\{ {{Z_0},{Z_1},{Z_2},...} \right\}} &\mathop {\lim \sup }\limits_{n \to \infty } \frac{{ {\mathbb E}\left[ {\sum\nolimits_{i = 0}^{n - 1} \beta  \left( {{Y_i},{Z_i},{Y_{i + 1}}} \right)} \right]}}{{ {\mathbb E}\left[ {\sum\nolimits_{i = 0}^{n - 1} {\left( {{Y_{i + 1}} + {Z_i}} \right)} } \right]}}\label{obj}\\
{\rm s.t.}~&{\rm{Prob}}\left\{  {\cal E}_j\le\mu\right\} \ge 1-\rho  ,\label{cons:1}\\
&Z_i\in\left[0,T\right],~\forall i,j \in \left\{1,2,...\right\}\label{cons:2},
\end{align}\label{p1}
\end{subequations}
with 
\begin{flalign}
\mu ={\mathop {\lim \inf }\limits_{n \to \infty } \frac{1}{n}{\mathbb E}\left[ {\sum\nolimits_{i = 0}^{n - 1} {{\psi _i}} } \right] }\label{cons:1A}
\end{flalign}
and$${\psi _i} = \int_{{S_i}}^{{S_i} + {Y_i} + {Z_i}} {\Phi \left( {P\left( t \right)} \right)dt}.$$}

\textcolor[rgb]{0,0,0}{The symbols $\sup$ in (\ref{obj}) and $\inf$ in (\ref{cons:1A}) are due to the fact $\left\{Y_1,Y_2,...\right\}$  {being} random variables and the result of the functions of $\left\{Y_1,Y_2,...\right\}$ {being different}  even with the given distribution. In (\ref{cons:1}), as it is hard to guarantee the harvested energy always being over a random number ${\cal E}_j$ in both mathematics and engineering, we use $\rho$ to denote the EH  outage probability such that the long-term average received energy is greater than the required energy with a probability no less than $1-\rho$. In (\ref{cons:2}), $T$ is the maximum waiting time.}


\section{Solution Approach}
Problem (\ref{p1}) is a non-convex problem due to the infinite number of probability constraints (\ref{cons:1})  and the objective function. In order to efficiently solve it, we first deal with (\ref{cons:1}) as follows.

 {Suppose the channel  between the sensor and the sink node follows Rayleigh distribution. Then,} the channel coefficient ${{{\left| {{h_2}} \left( {{S_j}} \right)\right|}^2}}$ follows the exponential distribution and its probability density function (PDF) can be expressed by
\begin{flalign}
{f_{{{\left| {{h_2}} \left( {{S_j}} \right)\right|}^2}}}\left( x \right) = \lambda {e^{ - \lambda x}}\label{Ray}
\end{flalign}
where $\lambda$ is the exponential distribution parameter. Then, following (\ref{Ray}), (\ref{cons:1}) can be rewritten as
\textcolor[rgb]{0,0,0}{\begin{flalign}
&{\rm{Prob}}\left\{ {{{\left| {{h_2}\left( {{S_j}} \right)} \right|}^2} \ge \frac{{\left( {{2^{\frac{C}{{{T_c}B}}}} - 1} \right)d_2^\alpha {T_c}B{n_0}}}{\mu}} \right\} \nonumber\\
&= 1 - \int_0^{\frac{{\left( {{2^{\frac{C}{{{T_c}B}}}} - 1} \right)d_2^\alpha {T_c}B{n_0}}}{\mu}} {{f_{{{\left| {{h_2\left( {{S_j}} \right)}} \right|}^2}}}\left( x \right)dx}\nonumber \\
& = \exp \left\{ { - \frac{{\lambda {\left( {{2^{\frac{C}{{{T_c}B}}}} - 1} \right)d_2^\alpha {T_c}B{n_0}} }}{\mu}} \right\} \ge 1-\rho. \label{cons:2A}
\end{flalign}}


 {By doing so, (\ref{cons:1})  with infinite number of constraints is equivalently transformed to be as  (\ref{cons:2A}) which is with only one constraint.}

Further, with the non-linear EH model, the output DC power of the sensor is limited by the maximum output DC power. That is, if the received signals are with relatively high power level, output DC power of the sensor  {will be} constant, i.e., the maximum output DC power. Therefore, it is assumed that the sensor is close to the power station and always works in the saturation state of the EH circuit. Then, ${\psi _i}$ can be given by
\begin{flalign}
{\psi _i} = \left({Y_i} + {Z_i}\right)M\label{auxiliary:1}.
\end{flalign}
By substituting (\ref{auxiliary:1}) into (\ref{cons:2A}), we have  that
\begin{flalign}
\mathop {\lim \inf }\limits_{n \to \infty } \frac{1}{n}{\mathbb E}\left[ {\left( {{Y_i} + {Z_i}} \right)} \right] \ge  \omega \label{cons:2B}
\end{flalign}
where $$\omega  =- \frac{{\lambda \left( {{2^{\frac{C}{{{T_c}B}}}} - 1} \right)d_2^\alpha {T_c}B{n_0}}}{{\ln \left( {1-\rho } \right)M}}.$$

 {
As illustrated in Figure \ref{AoI2}, $\beta_i$ is only relevant to ${Y_{i}}$, ${Z_{i}}$ and ${Y_{i+1}}$, which is consistent with (\ref{beta:B}).
Thus, we have the following remark.}
\newtheorem{Rem}{Remark}
\begin{Rem}
 {The optimal online transmission policy $Z^\star_i$ is only related to $Y_i$ not related to $\left\{Y_1,Y_2,...,Y_{i-1}\right\}$ and/or $\left\{Z_1,Z_2,...,Z_{i-1}\right\}$.\label{Remark:1}}
\end{Rem}
Intuitively, Remark \ref{Remark:1} indicates that the waiting time depends on the last transmission time and the distribution of the transmission time. The similar result can be found in \cite{AoI1} and \cite{AoI2} which shows that there exists an optimal \emph{stationary deterministic policy} with $Z_i={\cal Z}\left(Y_i\right)$.

As $Y_i$ and $Y_{i+1}$ are $i.i.d$, we have
\begin{flalign}
\mathop {\lim }\limits_{n \to \infty } \frac{1}{{n}}{\mathbb E}\left[ {\sum\nolimits_{i = 1}^{n - 1} \beta  \left( {{Y_i},{Z_i},{Y_{i + 1}}} \right)} \right] = {\mathbb E}\left[ {\beta \left( {Y,Z,{{Y'}}} \right)} \right]\label{numerator}
\end{flalign}
and
\begin{flalign}
\mathop {\lim }\limits_{n \to \infty } \frac{1}{{n}}{\mathbb E}\left[ {\sum\nolimits_{i = 0}^{n - 1} {\left( {{Y_{i + 1}} + {Z_i}} \right)} } \right] = {\mathbb E}\left[ {{Y'} + Z} \right].\label{denominator}
\end{flalign}

By submitting (\ref{numerator}) and (\ref{denominator}) into Problem (\ref{p1}), solving Problem (\ref{p1})  {is equivalently transformed into} the following problem.
\begin{subequations}
\begin{align}
&\mathop {\min }\limits_{ {\cal Z}\left(\cdot\right)} \frac{{{\mathbb E}\left[ {\beta \left( {Y,Z,{{{{Y'}}}}} \right)} \right]}}{{{\mathbb E}\left[ {Y + Z} \right]}}\label{p2:1}\\
&{\rm s.t.}~{\mathbb E}\left[ {{Y} + {Z}} \right] \ge  \omega\label{p2:cons:1}\\
&~~~~~Z\in\left[0,T\right]\label{p2:cons:2}
\end{align}\label{p2}
\end{subequations}

Problem (\ref{p2}) is an optimization problem with fractional objective function, which is still non-convex. Nevertheless, it is observed that the numerator of the objective function, i.e., ${{\mathbb E}\left[{\beta \left( {Y,Z,{{{{Y'}}}}} \right)}\right]}$, is convex in $Z$, and the denominator function, i.e., ${{\mathbb E}\left[Y'+Z\right]}$, is an affine function. Thus, fractional programming can be employed to handle Problem (\ref{p2}), and the main idea is summarized in the following Lemma 1.


\newtheorem{Lem}{Lemma}
\begin{Lem}
The optimal solution to Problem (\ref{p2}) can be achieved if and only if $\gamma^\star$ being the unique zero of the auxiliary function $F\left(\gamma\right)$ where
\begin{flalign}
&F\left( \gamma  \right) \buildrel \Delta \over = {{\mathbb E}\left[{\beta \left( {Y,Z,{{{{Y'}}}}} \right)}\right]} - \gamma  {{\mathbb E}\left[Y'+Z\right]}.\label{new.opt}
\end{flalign}\label{fractional_programming}
\end{Lem}
\begin{IEEEproof}
The proof of Lemma 1 is similar to that in \cite{FP,new5}, which is omitted here.
\end{IEEEproof}

Lemma 1 indicates that Problem (\ref{p2}) can be transformed into an equivalent problem where the original fractional objective function (\ref{p2:1}) is replaced by a function in subtractive form, i.e., (\ref{new.opt}). Thus, instead of solving Problem (\ref{p2}), we can solve the following auxiliary Problem (\ref{p4}) which has the same optimal solution to Problem (\ref{p2}).
\begin{flalign}
&\mathop {\min }\limits_{Z,\gamma}~ F\left( \gamma  \right)\label{p4} \\
&{\rm s.t.}~(\rm \ref{p2:cons:1}),(\rm \ref{p2:cons:2}).\nonumber
\end{flalign}

Problem (\ref{p4})  is still not jointly convex w.r.t. $Z$ and $\gamma$. But it is convex w.r.t. $Z$ (or $\gamma$) with a fixed $\gamma$ (or $Z$). Thus, we develop an algorithm as shown in Algorithm 1 to solve Problem (\ref{p4}), where the outer loop  is designed based on Dinkelbach's method. In particular, in the $q$-th iteration, with the given $\gamma$, i.e., $\gamma\left(q\right)$, by solving the following Problem (\ref{p5}),
\begin{flalign}
&\mathop {\min }\limits_{Z}~ F\left( \gamma\left(q\right)  \right)\label{p5} \\
&{\rm s.t.}~(\rm \ref{p2:cons:1}),(\rm \ref{p2:cons:2}),\label{p5:cons:1}\nonumber
\end{flalign}
the optimal ${Z}^\star\left(q\right)$ is derived, which can be calculated in terms of the following Theorem 1.


\newtheorem{The}{Theorem}
\begin{The}
The optimal solution to Problem (\ref{p5}) can be given by
\begin{equation}
Z^\star\left(q\right) = \left[ {\frac{1}{a}\ln \left( {\frac{{\gamma \left( q \right) + \eta  + 1}}{{{\mathbb E}\left[ {{e^{\left( {aY'} \right)}}} \right]{e^{\left( {ay} \right)}}}}} \right)} \right]_0^T.\label{Z}
\end{equation}
\end{The}

\begin{IEEEproof}
Since Problem (\ref{p5}) is convex, KKT conditions can be used  to optimally solve it. The Lagrangian of (\ref{p5}) can be given by (\ref{L}), where ${\eta}$ is the Lagrange multiplier associated with (\ref{p2:cons:1}) and ${f_Y}\left( y \right)$ is the PDF of $Y$. According to KKT conditions, the optimal $Z^\star\left(q\right)$ satisfies
\begin{flalign}
&\frac{{\partial \Gamma \left( Z^\star \right)}}{{\partial Z}} = \label{Z:2}
\\&~~~~ \left( {\left( {{e^{a\left( {y + {\cal Z}^\star(y) + Y'} \right)}}} \right) - 1 - \gamma\left(q\right)   - {\eta}} \right){f_Y}\left( y \right) = 0.\nonumber
\end{flalign}
Following (\ref{Z:2}), the optimal $Z^\star\left(q\right)$ can be calculated as  (\ref{Z}).
\end {IEEEproof}

\begin{Rem}
\textcolor[rgb]{0,0,0}{It is seen from Theorem 1 that for a given distribution of the transmission time, the waiting time decreases with the increment of the last transmission time.}
\end{Rem}

Note that $\eta$ is still unknown in (\ref{Z}). Thus, we employ a bisection method to find the optimal $\eta^\star$ in the inner loop  of Algorithm 1 where $\varepsilon$ denotes the small positive tolerance, $\iota$ is the lower bound of $\eta$ which can be set as $0$, and  $u$ is the upper bound of $\eta$ which can be set as a relatively large positive number. {Once the optimal $\gamma ^ \star$ and $\eta^\star$ are obtained, the corresponding $Z^\star$ is also derived which is the final \textcolor[rgb]{0,0,0}{numerical} solution to Problem (\ref{p2}).}

\begin{figure*}
\begin{flalign}
&{\mathcal L}\left(Z\right)={\mathbb E}\left[ {\frac{1}{a}\left( {{e^{a\left( {Y + Z + Y'} \right)}} - {e^{aY}}} \right) - Z - Y' - \gamma\left(q\right) \left( {Y' + Z} \right)} \right] + {\mathbb E}\left[ {{\eta}\left( {\omega  - Y - Z} \right)} \right] \label{L}  \\
&=\int_0^\infty  {\left( {{\mathbb E}\left[ {\frac{1}{a}\left( {{e^{a\left( {y +{\cal Z}(y) + Y'} \right)}} - {e^{aY}}} \right)} \right] - {\cal Z}(y) - {\mathbb E}\left[ {Y'} \right] - \gamma \left( q \right)\left( {{\mathbb E}\left[ {Y'} \right] + {\cal Z}(y)} \right) + {\eta}\left( {\omega  - y - {\cal Z}(y)} \right)} \right){f_Y}\left( y \right)dy} \nonumber\\
& \buildrel \Delta \over = \int_0^\infty  {\Gamma \left( Z \right){f_Y}\left( y \right)dy}  \nonumber
\end{flalign}
\hrule
\end{figure*}

\begin{algorithm}[t]\label{FP_GEE}
\caption{ Iterative solution approach based on Dinkelbach's Algorithm for solving Problem (\ref{p5})}
 \While{$F\left({\gamma\left(q\right)}\right)<\varepsilon$}{{\bf Initialize} ${\gamma\left(0\right)}$ with $F\left({\gamma\left(0\right)}\right)\ge0$\;
 Set $q=0$\;
 {\bf Initialize} $\iota \le{\eta}\le u$\;
\While{$u-\iota\ge\varepsilon$}{
 Update $\eta$=$\left(\iota+u\right)/2$\;
 Solving problem (\ref{p5}) by  calculating $Z\left(q\right)$ in terms of  (\ref{Z}) \;
 \eIf{ (\ref{p2:cons:1}) is not satisfied } {Update $\iota=\eta$\;}{Update $u=\eta$\;}
 }
 Update $$F\left( {\lambda \left( q \right)} \right)= {\mathbb E}\left[{{\beta \left( {Y,Z\left(q\right),{{{{Y'}}}}} \right)} }\right] - \gamma {\mathbb E}\left[Y'+Z\left(q\right)\right];$$\
Update
\begin{flalign}
\gamma \left( {q + 1} \right) = \frac{{\mathbb E}\left[{{\beta \left( {Y,Z\left(q\right),{{{{Y'}}}}} \right)} }\right]}{ {\mathbb E}\left[Y'+Z\left(q\right)\right]}\nonumber;
\end{flalign}\
 $q=q+1$\;}
 {\bf Return} $Z\left(q\right)$.
\end{algorithm}


\section{Simulation Results}\label{simulation}

This section represents some simulation results to show the validity and efficiency of the proposed \textcolor[rgb]{0,0,0}{solution approach} and present some interesting insights. First, we compare the proposed transmission policy with  two benchmark policies, i.e., zero time waiting policy and equal time waiting policy. Then, we show impact of the EH circuit saturation level  and EH outage probability on the system average U-AoI.

The simulation network scenario is shown in Figure \ref{Sys} and the simulation parameters are set according to \cite{Sim}. The distance between the sensor and the sink node is set as $2$m. The Rayleigh channel parameter $\lambda$ is set as $3$ and the pass loss factor $\alpha$ is set as $2$. The block length $T_c$ is set as $10^{-3}$s, the noise power spectral density is $-70$ dBm/Hz and the bandwidth $B$ is $10$MHz. The size of each update $C$ is set as $8$bits. For the non-linear EH model, we set $M$ as $20$mW which corresponds to the maximum output DC power, i.e., the charging power of the sensor. The EH outage probability $\rho$ is set as $1\%$.

 {We define an event $\left\{Y\le\kappa\right\}$ with a probability $\theta$, i.e. $\theta=\rm Prob\left\{Y\le\kappa\right\}$, \textcolor[rgb]{0,0,0}{which is used to characterize the congestion degree of the wireless networks}. Note that for a given $\theta$, the network gets more congested as $\kappa$ increasing. While for a given $\kappa$, the network gets less congested as $\theta$ increasing.}

\begin{figure}[t]
\begin{center}
\centerline{\includegraphics[ width=0.5\textwidth]{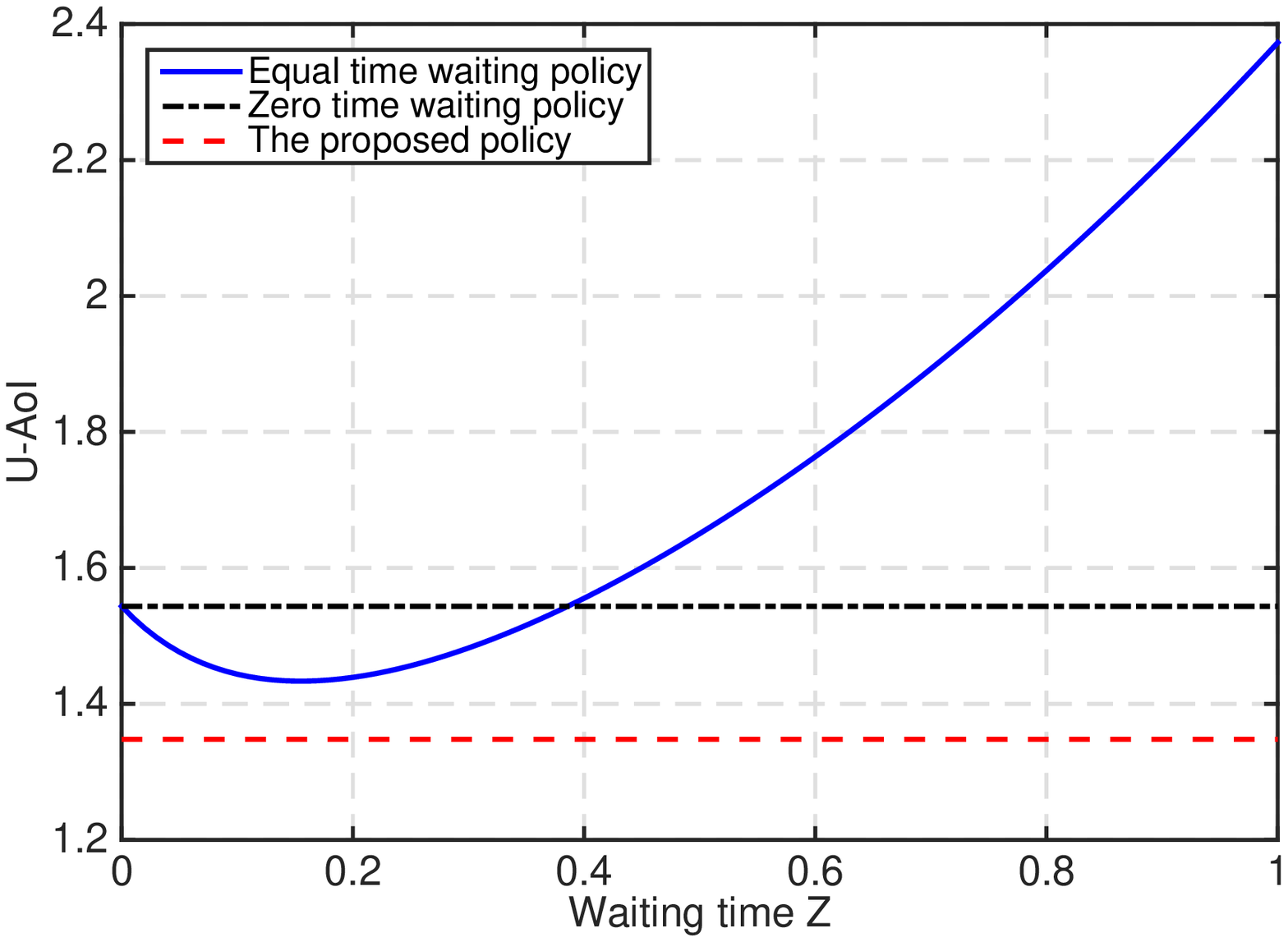}}
\caption{Average U-AoI v.s. waiting time with $\rm Prob\left\{Y\le0.1\right\}=0.7$.}
\label{prob7}
\end{center}
\end{figure}

\begin{figure}[t]
\begin{center}
\centerline{\includegraphics[ width=0.5\textwidth]{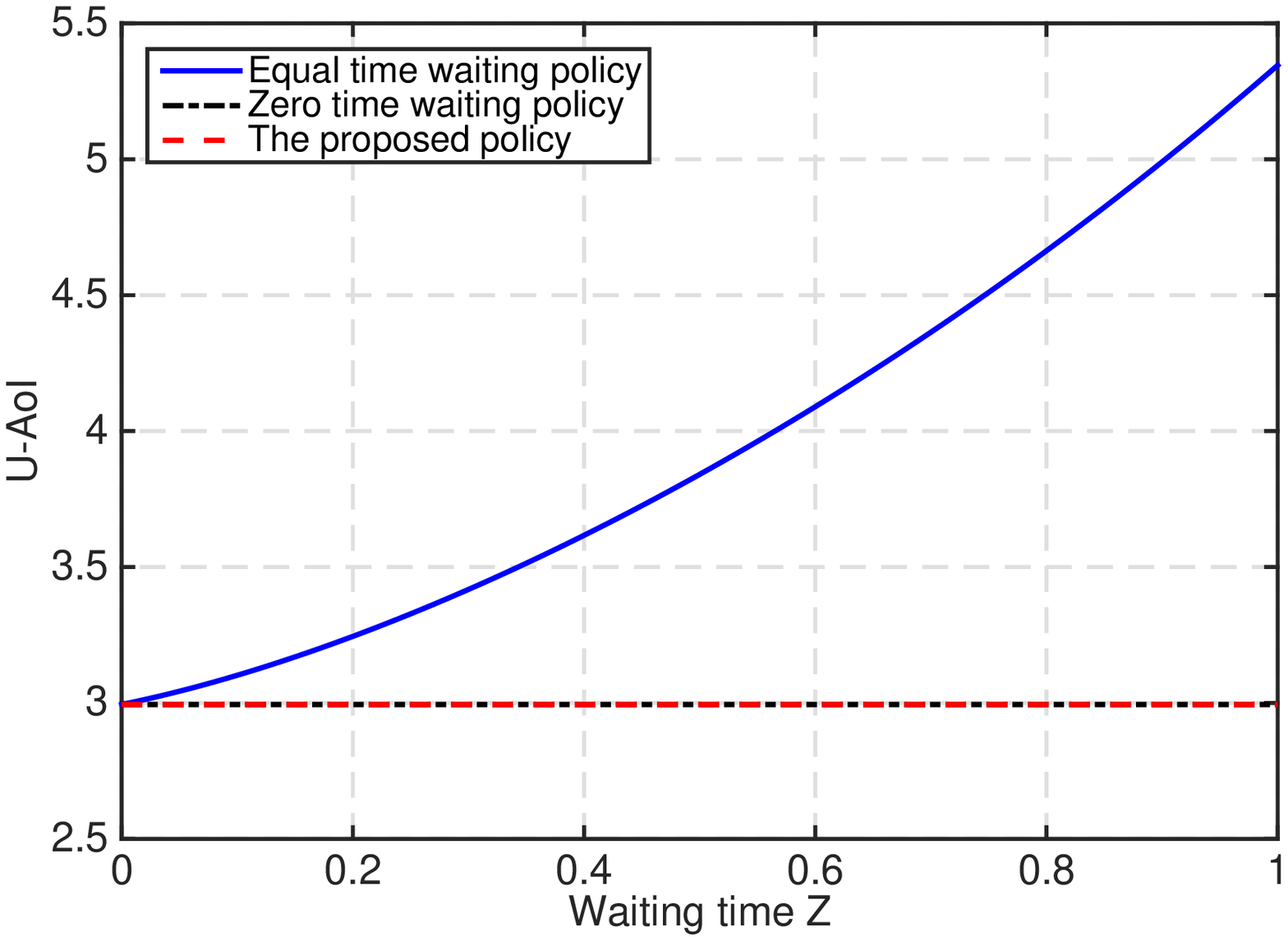}}
\caption{Average U-AoI v.s. waiting time with $\rm Prob\left\{Y\le0.1\right\}=0.3$.}
\label{prob3}
\end{center}
\end{figure}

Figure \ref{prob7} compares the proposed transmission policy with  {two benchmark policies when $M$ and $T$ are both set as $\inf$ so that the zero waiting policy can be always feasible, and $\theta=0.7$, i.e., $\rm Prob\left\{Y\le0.1\right\}=0.7$}. It is seen that the system average U-AoI first increases and then decreases with the  waiting time when the equal time waiting policy is adopted. The zero time waiting policy and the proposed policy do not change with the  waiting time. It is also observed that the proposed transmission policy is superior to the other two transmission policies in terms of the system average U-AoI. In Figure \ref{prob3}, we compare the mentioned three transmission policy when $M$ and $T$ are both set as $\inf$, and $\rm Prob\left\{Y\le0.1\right\}=0.3$. Different from Figure \ref{prob7}, it is seen that in this case, the system average U-AoI increases with the waiting time when the equal time waiting policy is adopted. Besides, the zero time waiting policy and the  equal time waiting policy are very close to the proposed policy.

With Figure \ref{prob7} and Figure \ref{prob3}, we note that even without EH constraint and maximum waiting time constraints, zero time waiting policy may not be optimal, especially when the network is not congested (e.g., $\rm Prob\left\{Y\le0.1\right\}=0.7$). It means that in this case, the sensor should be a little $lazy$. For instance, the transmission times of four consecutive updates are $\left\{0,0,0,2\right\}$. If zero time waiting policy is adopt, the average U-AoI  is $\frac{1}{{3}}\left({{{e^2} + e - 5}}\right)\approx {\rm{1.70}}$ when $a$ is set as $1$. However, if we insert a waiting time $0.9$ before transmitting the third update, the average U-AoI is $\frac{1}{{4}}\left({{{e^2} + 2e - 7}}\right)\approx {\rm{1.45}}.$ The reason is that transmitting the third update without waiting will waste the potential benefit of zero transmission time. Meanwhile, when the network is congested (e.g., $\rm Prob\left\{Y\le0.1\right\}=0.3$), zero time waiting policy is very close to the proposed transmission policy. In this case, the sensor should transmit updates without waiting. In fact, the proposed transmission policy adjusts the waiting time based on the transmission time (i.e., the network conditions), so that the sensor avoids high-frequent updates transmitting when the network is not congested and reduces waiting time when the network is congested. By doing so, the long-term system average U-AoI is therefore reduced.

\begin{figure}[t]
\begin{center}
\centerline{\includegraphics[ width=0.5\textwidth]{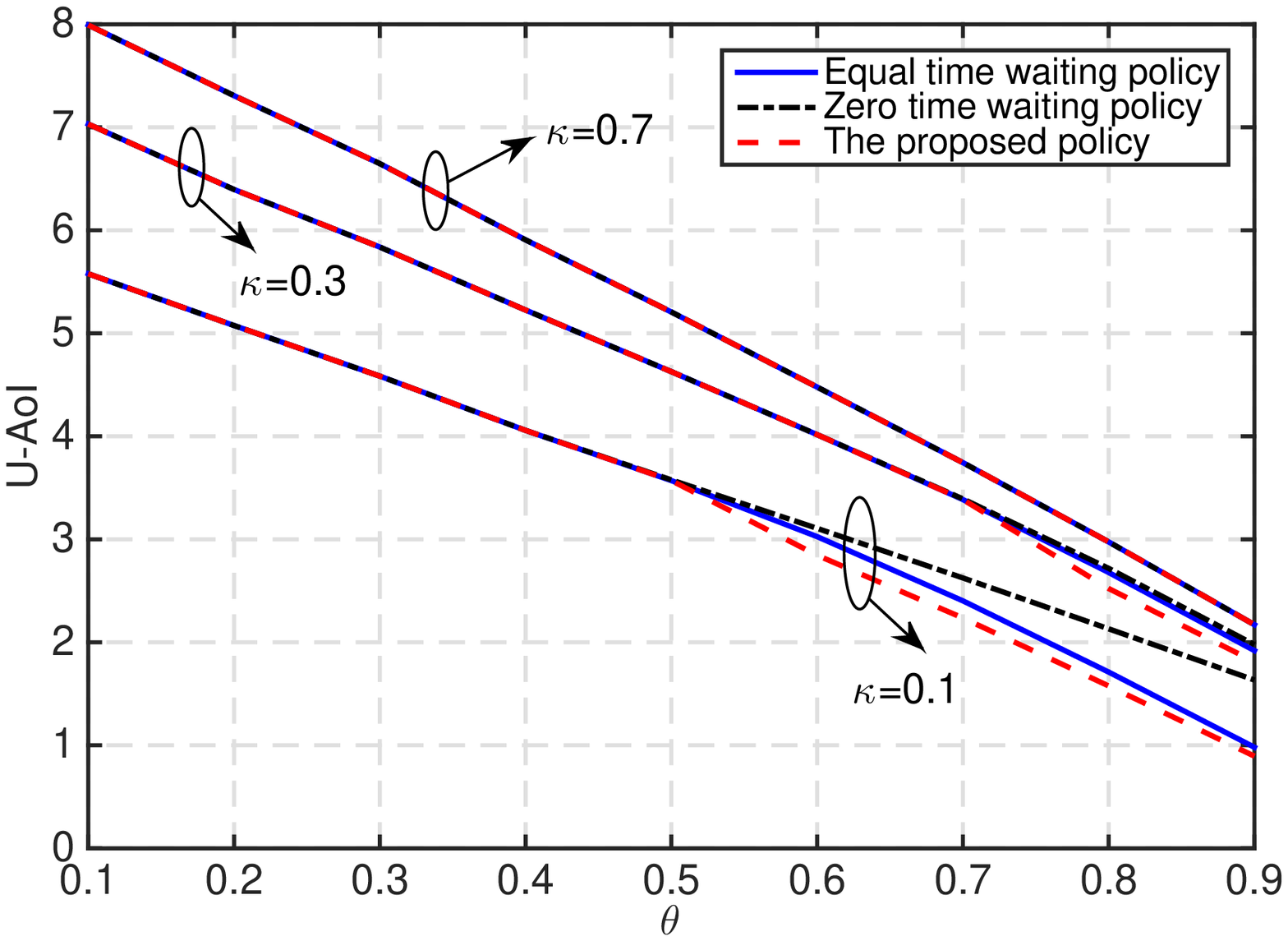}}
\caption{Average U-AoI v.s. $\theta=\rm Prob\left\{Y\le\kappa\right\}$.}
\label{prob}
\end{center}
\end{figure}

\textcolor[rgb]{0,0,0}{Figure \ref{prob} compares the mentioned three policies versus $\theta$ where $\kappa$ is set as $0.1$, $0.3$, and $0.7$, respectively. It is seen that the proposed transmission policy outperforms the other two policies as it adjusts the waiting time based on the transmission time. Besides, the equal time waiting policy is better than the zero time waiting policy because zero time waiting policy is actually a special case of equal time waiting policy. It is also observed that when $\kappa=0.3$ and $\kappa=0.7$, the proposed transmission policy is very close to the other two policies. However, when $\kappa=0.1$, the difference is relatively obvious, especially with a relatively large $\theta$. The reason is that when $\kappa$ is relatively large, the network is relatively congested. In this case, the sensor tends to transmit the next updates without waiting.  While, when $\kappa$ is relatively small, the network is relatively not congested. In this case, the sensor tends to wait a moment before transmitting next updates. The insight drawn based on  Figure \ref{prob} is consistent with that associated with Figure \ref{prob7} and Figure \ref{prob3}.}

\begin{figure}[t]
\begin{center}
\centerline{\includegraphics[ width=0.5\textwidth]{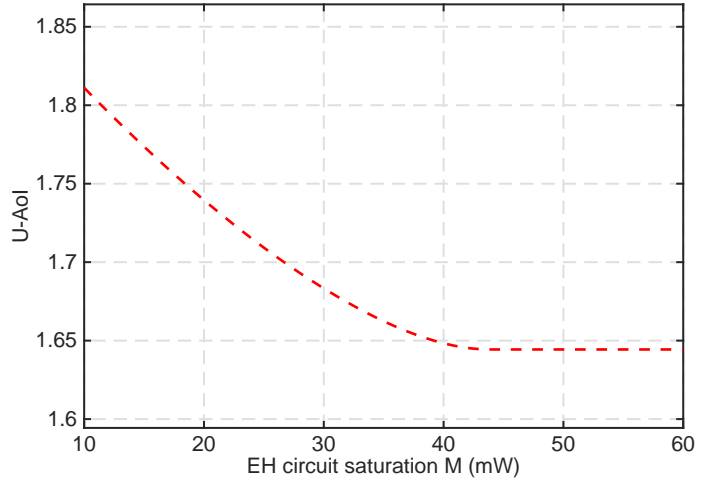}}
\caption{U-AoI v.s. EH circuit saturation  level.}
\label{AOIEH}
\end{center}
\end{figure}

\begin{figure}[t]
\begin{center}
\centerline{\includegraphics[ width=0.5\textwidth]{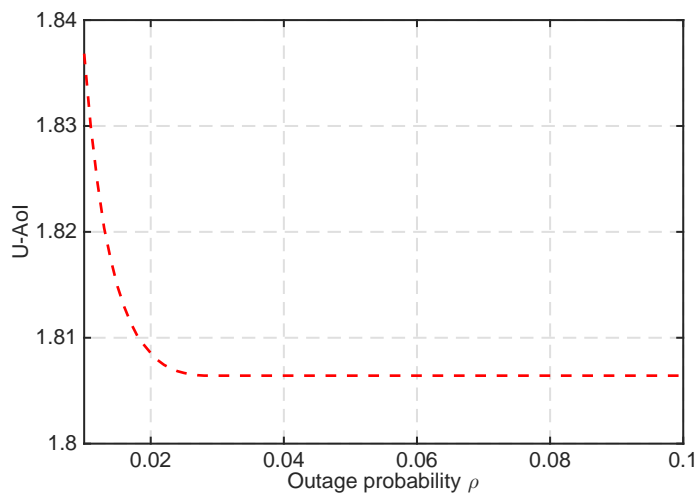}}
\caption{U-AoI v.s. outage probability.}
\label{outage}
\end{center}
\end{figure}

Figure \ref{AOIEH} shows the U-AoI versus the EH circuit saturation level, i.e., the maximum output DC power. It is seen that the average U-AoI first decreases and then keep unchanged with the EH circuit saturation. The reason is that when the output DC power is relatively low, the sensor need more time to charge itself for transmitting the next update. In this case, the waiting time may be larger than the optimal waiting time, and thus, the system average U-AoI is degraded. However, when the output DC power is relatively high, the required waiting time for charging can be lower than the optimal waiting time plus the last transmission time. Therefore, the system can always works with the minimal average U-AoI. Thus, for EH-enabled networks, to keep the information fresh, charging the sensor with high power level may cause the waste of energy.

\textcolor[rgb]{0,0,0}{The average U-AoI versus the outage probability is shown in Figure \ref{outage}. Similar to Figure \ref{AOIEH}, the average U-AoI first decreases and then keep unchanged with the outage probability. This is because the low outage probability requires high average EH power, which makes the energy causality constraint relatively hard to be satisfied.}

\section{Conclusion}\label{conclusion}
{This paper studied the AoI in EH-enabled networks where a sensor first scavenges energy from a power station and then, transmits its status update to a sink node. U-AoI was defined to capture thirst for the fresh update becoming more and more urgent as time elapsing. Then, an optimization problem was formulated to minimize the long-term average U-AoI  under constraint of energy causality. To solve the considered non-convex problem, we developed a two-layer  algorithm. Numerical results shows that the proposed transmission policy is superior to the zero time waiting policy and the equal time waiting policy in terms of U-AoI, especially when the networks are not congested. It is also found that sensor tends to transmit the next update without waiting when the network is congested while tends to wait a moment before transmitting the next update when the network is not congested. Besides, U-AoI first decreases and then keep unchanged with the charging power level and the EH outage probability.}

\begin{appendices}

\end{appendices}

\section*{Acknowledgment}
\addcontentsline{toc}{section}{Acknowledgment}
This work was supported in part by the National Key R\&D Program of China (No. 2016YFE0200900), in part by the General Program of the National Natural Science Foundation of China (NSFC) (No. 61671051), and also in part by the major projects of Beijing Municipal Science and Technology Commission  (No. Z181100003218010).

\end{document}